# The Rare Two-Dimensional Materials with Dirac Cones

*Jinying Wang, Shibin Deng, Zhongfan Liu, and Zhirong Liu*[*]

[1] Center for Nanochemstry, Colledge of Chemistry and Molecular Engineering, Peking University, Beijing 100871, China

[*]*Email address: LiuZhiRong@pku.edu.cn*

## ABSTRACT

Inspired by the great development of graphene, more and more works have been conducted to seek new two-dimensional (2D) materials with Dirac cones. Although 2D Dirac materials possess many novel properties and physics, they are rare compared with the numerous 2D materials. To provide explanation for the rarity of 2D Dirac materials as well as clues in searching for new Dirac systems, here we review the recent theoretical aspects of various 2D Dirac materials, including graphene, silicene, germanene, graphynes, several boron and carbon sheets, transition metal oxides $(VO_2)_n/(TiO_2)_m$ and $(CrO_2)_n/(TiO_2)_m$, organic and organometallic crystals, so-$MoS_2$, and artificial lattices (electron gases and ultracold atoms). Their structural and electronic properties are summarized. We also investigate how Dirac points emerge, move, and merge in these systems. The von Neumann-Wigner theorem is used to explain the scarcity of Dirac cones in 2D systems, which leads to rigorous requirements on the symmetry, parameters, Fermi level, and band overlap of materials to achieve Dirac cones. Connections between existence of Dirac cones and the structural features are also discussed.

## CONTENTS

    Organic and organometallic crystals

    Systems with a pseudospin of $S = 1$

    Artificial lattices: electron gases and ultracold atoms

**Why are Dirac materials so rare**

    Moving and merging of Dirac points

    Existent conditions of Dirac cones

    Structural features of the known 2D Dirac materials

**Conclusions**

# INTRODUCTION

Two-dimensional (2D) crystal was thought to be unstable in nature for a long time until 2004 when graphene, a one-atom-thick honeycomb structure composed of carbon atoms, was successfully prepared[1]. The discovery of graphene has attracted great interest because of the promising prospects of graphene in both basic and applied research[2-3]. In particular, the Dirac-cone structure gives graphene massless fermions, leading to half-integer[4-5]/fractional[6-7]/fractal[8-10] quantum Hall effects (QHE), ultrahigh carrier mobility[11], and many other novel phenomena and properties[12-13]. The great development of graphene also opens a door for discovering more 2D materials[14-15].

Up to now, over hundreds of 2D materials have been found, including group-IV compounds, binary systems of group III-V elements, metal chalcogenides, complex oxides, and so on[14-15]. But among them, only graphene[4-5, 16], silicene and germanene (graphene-like silicon and germanium, respectively)[17], several graphynes (sp-$sp^2$ carbon allotropes)[18-19], and some other systems[20-31] have been predicted to be Dirac materials (summarized in Fig. 1). Moreover, only Dirac cones in graphene have been truly confirmed experimentally[4-5]. Why are the 2D Dirac materials so rare? What are the special structure and properties for them? And how can we find new massless Dirac systems?

This paper reviews the theoretical studies on the 2D materials with Dirac cones. We start with a survey on the known 2D Dirac materials up to date, where emphases are put on their structural and electronic properties. Then we discuss the underlying mechanism for the rarity of 2D Dirac systems, i.e., how Dirac cones emerge and merge in a 2D crystal, and what conditions are required for a 2D Dirac system.

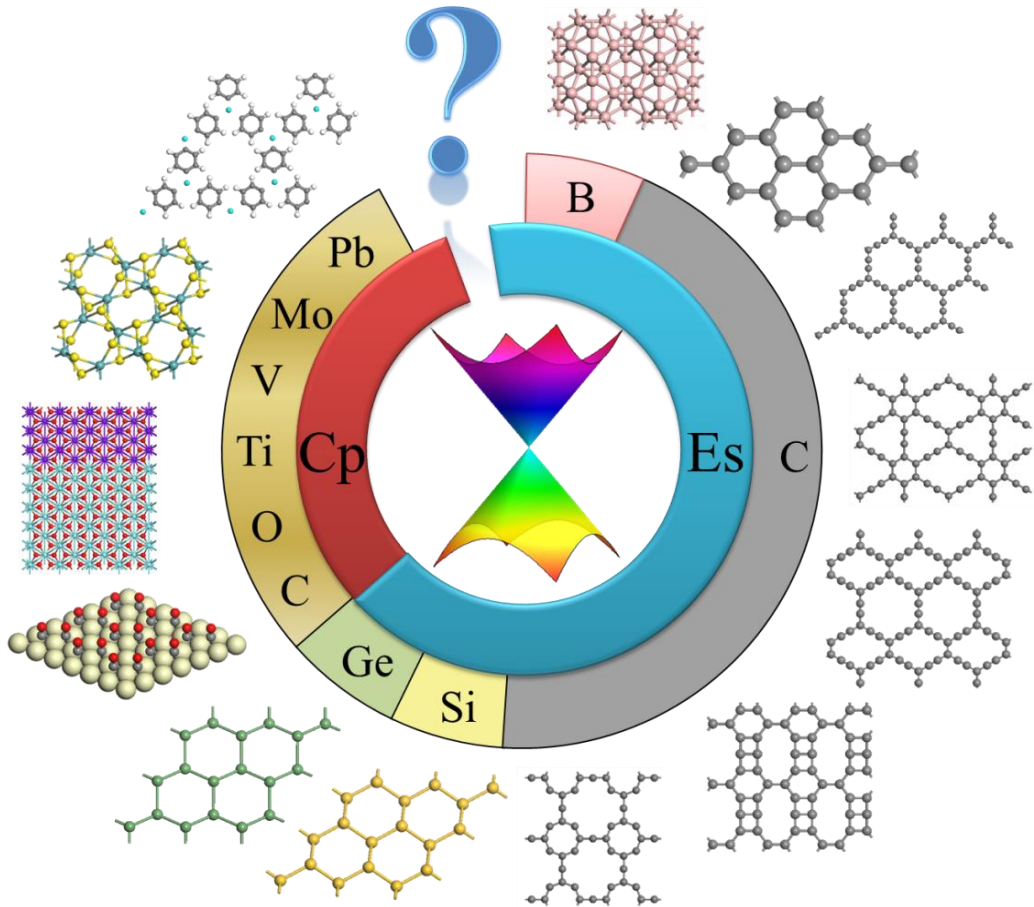

Figure 1. 2D materials with Dirac cones, which contain various elementary substances (Es) and compounds (Cp) with the elements of B, C, O, Si, Ge, Mo, V, Pb, etc. In detail, there are *Pmmn* boron, graphene, $\alpha$-graphyne, 6,6,12-graphyne, 14,14,18-graphyne, S-graphene, square carbon SG-10b, silicene, germanene, CO on Cu(111), $(VO_2)_3/(TiO_2)_5$, so-$MoS_2$, and $Pb_2(C_6H_4)_3$ in the clockwise direction.

## The known 2D Dirac materials

### Graphene

Graphene is the most typical 2D material with Dirac cones[12]. It can be regarded as a single layer of graphite, which has two C atoms per unit cell arranged in a hexagonal lattice (Fig. 2a). The C-C bond length in graphene is 1.42 Å, and the lattice constant is 2.46 Å. Graphene is rather stable because the C atoms are $sp^2$-hybridized and bind together with both σ and π bonds.

The electronic structure of graphene can be described by a tight-binding (TB) approach[12, 16]. Under nearest-neighbor approximation, the Hamiltonian is simplified into a $2 \times 2$ matrix for each wave vector **k**:

$$H(\mathbf{k}) = \begin{bmatrix} \varepsilon_0 & -\sum_{i=1,2,3} t_i \exp(i\mathbf{k} \cdot \mathbf{d}_i) \\ \text{c.c.} & \varepsilon_0 \end{bmatrix} \quad (1)$$

where $\mathbf{d}_i$ ($i$ = 1, 2, 3) are vectors that connect a C atom to its three nearest neighbors, $t_i$ are the corresponding hopping energies, and $\varepsilon_0$ is the on-site energy. The energy bands are thus solved to be

$$E_\pm(\mathbf{k}) = \pm \left| \sum_{i=1,2,3} t_i \exp(i\mathbf{k} \cdot \mathbf{d}_i) \right| \quad (2)$$

by setting the Feimi level to be $\varepsilon_0 = 0$. For the equilibrium structure, $t_i \equiv t_0$ (≈ 2.7 eV) and $\mathbf{d}_i \equiv r_0$ (≈ 1.42 Å), and the valence and conduction bands contact at $\mathbf{K}$ and $\mathbf{K}'$ points of the hexagonal Brillouin zone. Expanding the energy bands around $\mathbf{K}$ (or $\mathbf{K}'$) gives:

$$E_\pm(\mathbf{q}) = \pm \hbar \upsilon_F |\mathbf{q}| \quad (3)$$

where $\mathbf{k} = \mathbf{K} + \mathbf{q}$, and $\upsilon_F = 3 t_0 r_0 / 2\hbar$ is the fermi velocity (~$10^6$ m/s). Eq. (3) shows that graphene has a cone-like band structure with linear dispersion near $\mathbf{K}$ (or $\mathbf{K}'$) points, similar to a relativistic particle. The TB results are consistent with first-principles calculations at low energy regions (Fig. 2b)[32]. The density of states (DOS) per unit cell (with a degeneracy of 4 included) near Fermi level is expressed as

$$\rho(E) = \frac{4|E|}{\sqrt{3}\pi t_0^2} \quad (4)$$

Thus, graphene is a gapless semiconductor with zero DOS at Fermi level. The Hamiltonian near $\mathbf{K}$ (or $\mathbf{K}'$) points can be also transformed[12, 33] into:

$$H(\mathbf{q}) = \upsilon_F \mathbf{p} \cdot \boldsymbol{\sigma} \quad (5)$$

where $\mathbf{p} = -i\hbar \nabla$ is the momentum operator, and $\boldsymbol{\sigma}$ is the Pauli matrice. Eq. (5) is identical to the massless Dirac equation (or Dirac-Weyl equation with spin $S = 1/2$) by replacing $c$ (speed of light) with $\upsilon_F$. Therefore, the $\mathbf{K}$ (or $\mathbf{K}'$) points are also called Dirac points, and the linear band structure is named as Dirac cone. Graphene possess massless Dirac fermions with pseudospins of ±1/2.

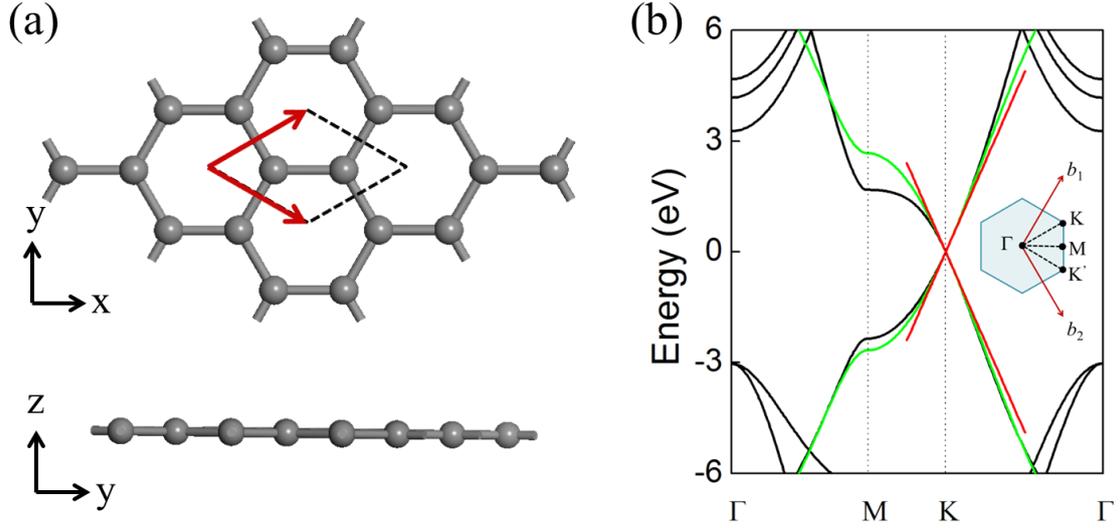

Figure 2. (a) Atomic structure and (b) energy bands of graphene. The energy bands were calculated by first-principles[32] (black line) and TB model with Eq. (2) (green line) and Eq. (3) (red line).

Many novel physical phenomena and electronic properties have been found in graphene due to the Dirac-cone structure. For example, a particular Landau level form in graphene when a uniform perpendicular magnetic field $B$ is applied[5, 34],

$$E(N) = \text{sgn}(N)\sqrt{2e\hbar v_F^2 |N| B} \tag{6}$$

while $E(N) \propto N$ in normal semiconductors. The Landau levels in 2D systems can be verified by Shubnikov-de Haas oscillations or QHE (if the magnetic field is strong enough). Different from the conventional 2D systems, graphene shows a half-integer QHE[4-5] with

$$\sigma_{xy} = (4e^2/h)(N+1/2) \tag{7}$$

which can be explained by the pseuospins and degeneracy of Dirac fermions. When interactions between electrons and magnetic flux quanta become very strong, new quasiparticles with fractional electronic charge can form and lead to correlated electron phenomena. Fractional QHE has been observed in suspended graphene[6-7] and graphene on $h$-BN[35] by reducing external perturbations, and the conductance $G$ can be expressed as

$$G = v e^2/h \tag{8}$$

where $v = p/q$ is the filling factor with $p$ and $q$ integers. Phase transitions between different fractional quantum Hall states have also been observed in suspended graphene, suggesting changes in spin and /or valley polarization[36]. In 1976, Hofstadter predicted a recursive electronic structure for 2D-confined electrons in magnetic field, which is called Hofstadter's butterfly[37]. This physical picture has been verified by the observation of fractal QHE in monolayer[9-10] and bilayer[8] graphene on *h*-BN when the magnetic length is comparable to the size of the superlattice.

Besides for the various QHEs, ultrahigh carrier mobility has also been found in graphene due to the massless Dirac-cone structure[13]. For graphene on Si/SiO$_2$ substrate, charge impurities are main scattering source which limit the mobility to be the order of $10^4$ cm$^2$ V$^{-1}$ s$^{-1}$ at low temperature[38]. The mobility of graphene on *h*-BN can be $\sim 6 \times 10^4$ cm$^2$ V$^{-1}$ s$^{-1}$ which is three times larger than that on Si/SiO$_2$, because the former substrate is flatter and has less charge impurities[39]. If all extrinsic scatterings are excluded, graphene has intrinsic mobility of $2 \times 10^5$ cm$^2$ V$^{-1}$ s$^{-1}$ at room temperature[40-41]. Castro et al. have studied the scattering mechanism of suspended graphene at different temperatures[42]. They found that flexural phonons limit the intrinsic mobility for temperature higher than 10 K. Various electron-phonon couplings have been investigated in a recent work which emphasizes the contribution of high-energy, optical, and zone-boundary phonons to electrical resistivity at room temperature[43]. Li et al. has also revealed that both longitudinal acoustic (LA) and transverse acoustic (TA) phonons are important in determining the intrinsic mobility of graphene[44].

The electronic properties of graphene are affected by many factors in reality, like ripples and substrates. Suspended graphene shows intrinsic ripples with height of ~1 nm and size of ~10 nm due to the thermal fluctuation[45-46]. Midgap states, nonzero DOS at fermi energy, and charge inhomogenity (electron-hole puddles) can form due to the ripple-induced modulation of hopping terms and localized states[47]. Graphene on a substrate also exhibits corrugations because of the substrate roughness and the lattice mismatch. For example, the typical corrugation is 0.2 nm in height for graphene on SiO$_2$ but only 0.02 nm for graphene on mica, which is in agreement with the substrate morphology[48]. For graphene on substrates, the charge inhomogenity mainly arises from substrate-induced charge impurites[49] and can be improved by using *h*-BN instead of SiO$_{2}$[50]. The interaction between graphene and substrates also contains strain effect, charge transfer, orbital hybridization, etc. A large uniaxial strain up to 24% might open the bandgap of graphene, but the Dirac cones keep robust for small and moderate uniform deformations[51]. The uniaxal and shear strains cause a moving of Dirac points[51-52] and anisotropic Fermi surface for graphene[51, 53]. Pereira and Castro Neto have proposed strain engineering of graphene by patterning the substrates[54]. The local-strain-induced gauge fields can be tailored to generate tunnel suppression, transport gap, and electron confinement in graphene[54]. Especially for a local triangular strain in graphene, the gauge field acts as a pesudomagnetic field which satisfies a simple model derived by Guinea et al.[55],

$$B_s = 8\beta c / a \tag{9}$$

where $\alpha$ is the lattice constant, $\beta = -\partial \ln t / \partial a \approx 2$ (*t* is hopping energy), $c = \Delta_m / D$ is the largest strain $\Delta_m$ for a disc of diameter *D*. When $\Delta_m$ = 10 % and *D* = 100 nm, $B_s$ is estimated to be

~40 T and the largest Landau gap ~0.25 eV, resulting an observable quantum hall effect[55]. Landau levels have been detected in graphene nanobubbles with a strain-induced pesudomagnetic field greater than 300 T[56]. The graphene-metal interfaces can be divided into two classes according to the binding energies $E_b$[57]. Grapene on Al, Ag, Cu, Au, and Pt surfaces which has weak adsorption (physisorption, $E_b < 0.5$ eV) preserves the Dirac-cone structure and show n- or p-doping because of charge transfer. However, graphene interacts strongly with Co, Ni, Pd, and Ti (chemsorption, $E_b > 0.8$ eV) by p-d hybridization, leading to an opened band gap of graphene submerged under the conducting states from metals[57].

**Silicene and Germanene**

The great success on graphene has inspired people to search for other 2D Dirac materials. Since carbon belongs to group IVA, silicon and germanium in the same group attract more and more attention. The critical problems are whether graphene-like silicon (silicene) and germanium (germanene) exist and what their electronic properties are.

Using first-principles calculations, Ciraci et al.[17] have predicted that silicene and germanene have stable low-buckled honeycomb structures (Fig. 3a and 3b). The bond lengths in silicene and germanene are 2.25 and 2.38 Å, respectively, which are much longer than the C-C length in graphene. The enlarged bond lengths weaken the π-π interaction and cause distinct coupling of σ and π bonds to form buckled structures. The buckling distance is 0.44 Å for silicene and 0.64 Å for germanene. Both low-buckled silicene and germanene are semimetals with Dirac cones at **K** points (Fig. 3c and 3d), and the Fermi velocity was estimated to be ~$10^6$ m/s, very close to that of graphene.

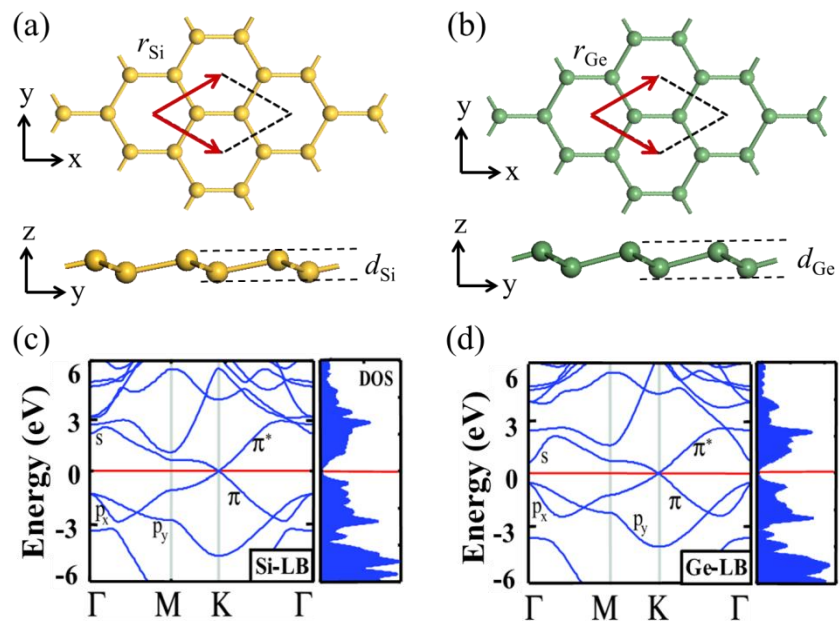

Figure 3. Atomic and electronic structures of (a, c) silicene and (b, d) germanene [17] (Copyright 2009 The American Physical Society).

However, the buckled structures greatly affect the electronic properties of silicene and germanene and bring about new physics beyond graphene. Liu et al.[58] have reported that the spin-orbital coupling (SOC) increases with the buckling degree and opens a band gap of 1.55 meV in silicene and 23.9 meV in germanene at Dirac points (Fig. 4a). The energy dispersion near Dirac points is written as:

$$E(\mathbf{k}) = \pm\sqrt{(v_F \hbar \mathbf{k})^2 + \Delta_{so}^2} \quad (10)$$

where $\Delta_{so}$ is the effective SOC. The spin-orbit band gap and nontrivial topological properties might result in detectable quantum spin Hall effect (QSHE). Considering the buckled structures, Ni et al.[59] have proposed to open the band gap of silicene or germanene by applying a vertical electric field $E_z$ to break the atomic symmetry (Fig. 4b). Drumond et al.[60] have considered the effect of SOC and vertical electric field simultaneously and have derived

$$E_{\sigma\xi,\pm}(\mathbf{k}) = \pm\sqrt{(\hbar v_F \mathbf{k})^2 + (\Delta_{so} + \sigma\xi\Delta_z)^2} \quad (11)$$

where $\xi = \pm 1$ distinguishes $\mathbf{K}$ and $\mathbf{K}'$, $\sigma = \pm 1$ distinguishes spin up and down, and $\Delta_z$ is a half of the electric-field induced band gap. Thus, silicene becomes a semimetal at a critical field $E_c \approx 20$ mV Å$^{-1}$ when $\Delta_{so} = \pm\Delta_z$. And it undergoes a transition from a topological insulator (TI) when $|E_z| < E_c$ to a simple band insulator (BI) when $|E_z| > E_c$. Similar results (Fig. 4c) have been obtained by Ezawa with further consideration of Rashba SOC, and a coexistence of TI, BI and metal regions in a silicene sheet by application of an inhomogeneous electric field has been proposed[61]. Ezawa also depicted two phase diagrams of silicene with exchange field $M$ and electric field $E_z$ and with light $A$ and electric field $E_z$, which include various insulating, semimetal and metallic states (Figs. 4d, e)[62-63]. Recently, a new quantum state, valley-polarized quantum anomalous Hall state (Fig. 4f) has been predicted in silicene through tuning the Rashba SOC[64]. The intrinsic carrier mobilities of silicene, germanene and graphene have the same order of magnitude under deformation potential theory (only the electron-acoustic phonon interaction is considered)[65-66], but the electron-optical phonon coupling in silicene or germanene is ~25 times smaller than that in graphene[67]. Based on the state transition of silicene at electric fields, a silicene-based spin filter has been proposed and demonstrated to exhibit 98% spin polarization[68], and a thermal-induced pure valley and spin currents has also been found in silicene junctions[69].

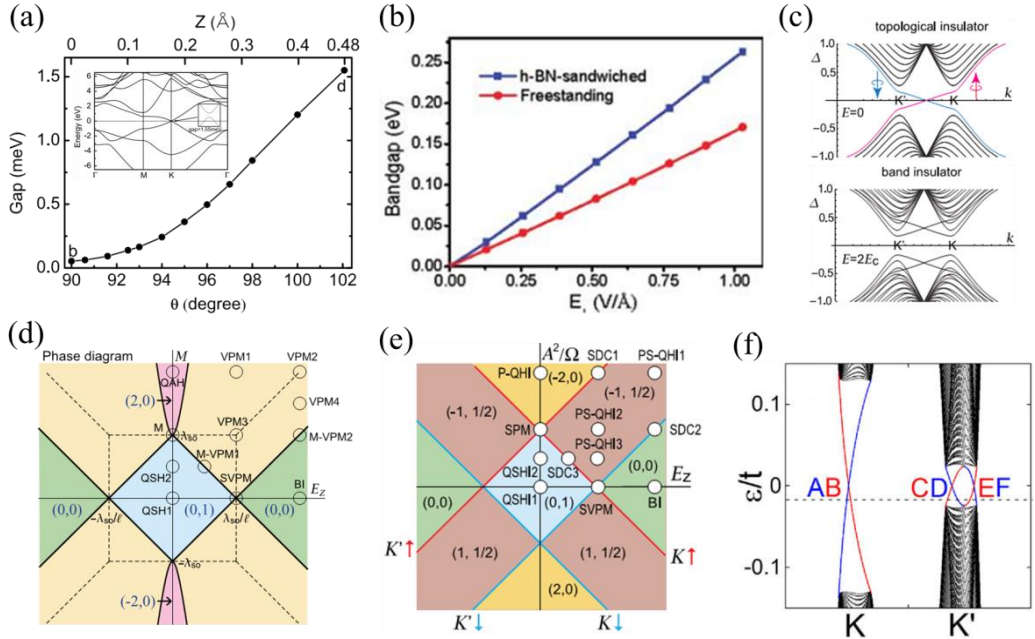

Figure 4. (a) The evolution of SOC-induced gap with buckling for silicene[58] (Copyright 2011 American Physical Society). (b) The band gap of silicene as a function of vertical electric field[59] (Copyright 2011 American Chemical Society). (c) the band structures of a silicene nanoribbon when $E_z = 0$ and $E_z = 2E_c$ [61]. (d) Topological phase diagram of silicene in $E_z - M$ plane[62] (Copyright 2012 American Physical Society). (e) Topological phase diagram of silicene in $E_z - A$ plane[63] (Copyright 2013 American Physical Society). (f) The electronic structure of a zigzag-terminated silicene with the valley-polarized QAHE[64] (Copyright 2014 American Physical Society).

The predicted novel properties make silicene and germanene promising in electronics, but can silicene or germanene be obtained experimentally? Aufray et al.[70] have reported an epitaxial growth of a silicene sheet on Ag(111). The STM and LEED images show a (2√3×2√3)R30° superstructure (with respect to Ag), and the Si-Si distance is 0.18~0.19 nm. Later studies revealed that silicene on Ag(111) has various superstructures[71-74], i.e. (√3×√3)R30°-Si (with respect to Si), (4×4), (√13×√13)R13.9°, and (√7×√7) R19.1 ° phases, etc. The superstructures are sensitive to the silicon coverage and substrate temperature[71-72], and the Si-Si distance varies between 2.28 and 2.5 Å[74]. Although several groups declared that they observed Dirac-cone structure of silicene on Ag(111)[75-76], the conclusion is controversial[77-78], especially after Takagi et al.[78] showed that the Dirac cones disappear for (4×4) silicene on Ag(111). Takagi et al. suggested that the linear dispersion of silicene on Ag(111) measured by Angular-resolved photoelectron spectroscopy (ARPES)[76] originate from the Ag bulk sp-band[78]. More and more works support the absence of Dirac cone in silicene on Ag(111) due to the strong interaction between Si and Ag and symmetry breaking[79-83]. Silicene has also been successfully grown on $ZrB_2(0001)$[84], Ir(111)[85], and MoS$_2$[86]. Silicene on ZrB$_2$(0001) is a (√3×√3)-reconstructed Si honeycomb sheet containing three kinds of Si atoms, which has a direct band gap[84]. For silicene on Ir(111), a buckled (√3×√3)-Si sheet with an undulation is formed[85]. Quhe et al. predicted that the absence of Dirac cones for silicene on various metal substrates is common because of the strong band hybridizations[87]. But the Dirac cones might be restored by intercalating alkali metal atoms between silicene and substrates[87]. Recently, a highly buckled silicene nanosheet has been reported to form on MoS$_2$[86]. First-principles calculations have shown that silicenes on MoS$_2$, MoSe2 and GaTe are metallic while those on MoTe$_2$, GaS, and GaSe are gapless semiconductors[88]. Theorists has also proposed several substrates with weak silicene-substrate interaction to reserve the Dirac fermions, like *h*-BN and hydrogen-processed Si(111)[80-81, 89]. Up to now, no freestanding silicene has been obtained, and the Dirac cone of silicene has not been confirmed. A new substrate is needed for both growing silicene and preserving the Dirac cones.

Since germanene has similar electronic properties to silicene, the growth of germanene is also attractive. In 2014, Gao et al. first reported the fabrication of germanene sheet on Pt(111) which has a distorted (√19×√19) superstructure with respect to the substrate[90]. Calculations have shown that germanene on Ag(111)/*h*-BN has an adsorption energy of -464~-428/-130 meV per Ge atom and is metallic/semiconducting[91]. Germanene on MoS$_2$ has also been discussed theoretically which is a p-doped semiconductor with a band gap of 24 meV[92].

More honeycomb sheet of group IVA elements and III-V binary compounds have been investigated theoretically to give semiconducting properties probably except SiGe[93-94]. Stanene which is a honeycomb sheet of tin atoms has been predicted to be a sizable-gap QSH insulator[95] but might be too unstable to exist really.

**Graphynes**

Graphynes are a series of 2D carbon allotropes composed of sp and sp$^2$ hybridized atoms, which were proposed by Baughman et al. in 1987[96]. Graphynes have various structures based on the ratio and arrangement of sp and sp$^2$ C atoms and possess high thermal stability[96]. The C-C bonds in most graphynes fall into three types: the triple bond (C≡C) with a length of 1.21~1.24 Å, the double bond (C=C) with a length of 1.31~1.35 Å, and the conjugate double bond (C…C) with a length of 1.39~1.45 Å[97-98]. Early first-principles calculations showed that graphynes can be semiconductors or semimetals or metals[97-99]. In 2012, D. Malko et al.[18] reported that *α*-, *β*-, and 6,6,12-graphyne (Fig.5a-c) could have Dirac-cone structures (Fig.5d-f). Both *α*-graphyne and *β*-graphyne exhibit hexagonal symmetry, but the Dirac points in the former are at **K** points while they locate along **Γ**-**M** line in the latter. In addition, 6,6,12-graphyne has a rectangular lattice and possesses two kinds of distorted Dirac cones with self-doping. These results not only indicate that graphynes might have more versatile properties than graphene but also break the stereotype that a honeycomb or hexagonal structure is a prerequisite for Dirac cones.

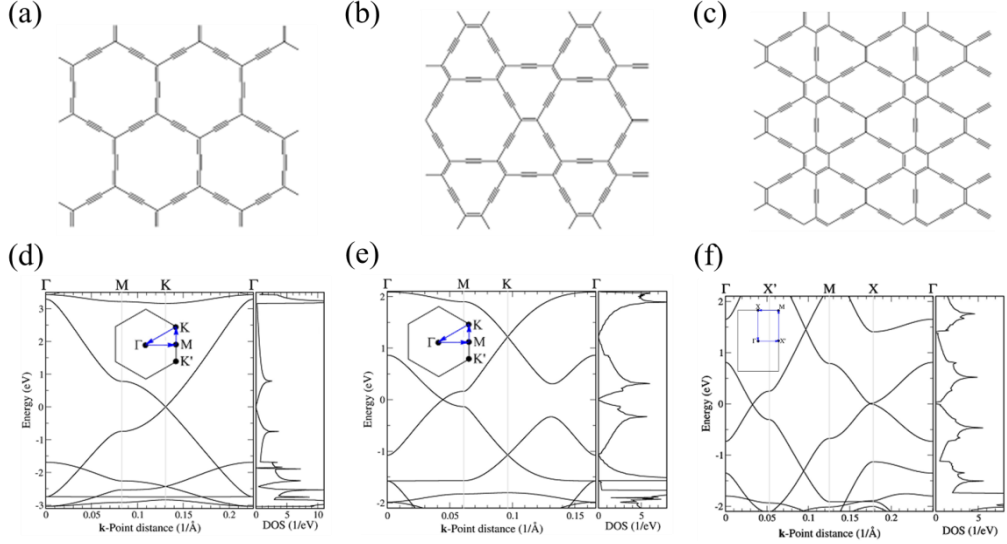

Figure 5. Atomic and band structures of (a,d) *α*-graphyne, (b,e) *β*-graphyne, and (c,f) 6,6,12-graphyne[18] (Copyright 2012 American Physical Society).

Why are some graphynes semimetals with Dirac cones while others are semiconductors? Kim et al.[100] have first noticed that *α*-, *β*-, and *γ*-graphyne are topologically equivalent to graphene where the role of triple bonds can be simplified as effective hopping terms. Liu et al.[101] have also used a TB model with effective hopping terms to fit the band structures of a few graphynes very well. Huang et al. then derived a criterion for the existence of Dirac cones in graphynes[19],

$$\frac{t_r}{t_b}=1 \quad \text{or} \quad -2\leq \frac{t_r}{t_b} \leq -1 \qquad (12)$$

where $t_r$ and $t_b$ are the two kinds of effective hopping energies in a graphene-like TB model to describe the band structures of graphynes. Eq. (12) indicated that whether Dirac cones exist or not in graphynes is determined by the combination of hopping energies. With $t_r$ and $t_b$ values derived from first-principles calculations, this criterion successfully explains the properties of various graphynes, including *α*-, *β*-, *γ*- and 6,6,12-graphyne. It also suggests that more graphynes with Dirac cones can emerge by tailoring the hopping terms. Actually, they have predicted that two graphynes (14,14,14-graphyne and 14,14,18-graphyne) without hexagonal symmetry have Dirac cones[19]. Three other rectangular graphynes have recently been designed, and two of them (6,6,18-graphyne and *h*-12,12,20-graphyne) have anisotropic Dirac cones along $\Gamma$-$X'$ lines[102]. Hexagonal *δ*-graphyne which is more stable than *α*- or *β*-graphyne shows Dirac-cone structure, too[103]. More importantly, *δ*-graphyne becomes a topological insulator under SOC with an induced gap of 0.59 meV and a $Z_2$ topological invariant of $v=1$ [103]. A TB approach to investigate SOC effect in more graphynes has been developed, and rich phase transitions were predicted[104-105]. Besides designing new graphynes, another way to construct Dirac systems is tuning the structure of known graphynes. Cui et al. have observed a transition from semiconductor to semimetal with Dirac cones in graphdiyne under uniaxial tensile strain[106]. The reverse way is also possible, e. g., a transition from gapless to finite gap system has also been achieved by an application of tensile strain to 6,6,12-graphyne[107].

The room-temperature intrinsic carrier mobilities of $\alpha$-, $\beta$-, and 6,6,12-graphyne calculated under deformation potential theory vary from $0.8 \times 10^4$ to $5.4 \times 10^5$ cm$^2$ V$^{-1}$ s$^{-1}$, which can be even higher than that of graphene[108]. Further consideration of various electron-phonon interactions has indicated that the mobilities of $\alpha$- and $\gamma$-graphynes are limited by LA phonon scatterings when temperatures are less than 600 K and are ~$10^4$ cm$^2$ V$^{-1}$ s$^{-1}$ at room temperature[109]. Recently, an analytic formula for the intrinsic mobility of 2D Dirac systems has been derived by Li et al. with considering the influences of LA and TA phonon modes and anisotropy. It reveals that the high mobility of graphynes originate from a suppression of TA phonon scatterings[44]. Directional transport properties of 6,6,12-graphyne have been predicted, which can be further manipulated by strain[110].

Although novel electronic properties have been proposed in graphynes, the experimental progress is still in its infancy[111]. Among the various graphynes, only graphdiyne-like films have been reported to be synthesized by homocoupling of monomers[112]. The related reaction mechanisms have been investigated both experimentally and theoretically[113-115], and the main difficulty to grow graphynes is how to reduce the side-reactions.

**Rectangular carbon and boron allotropes**

Carbon is the element with the most allotropes. More 2D carbon lattices with Dirac cones are expected after the discovery of graphene and graphynes. Liu et al. proposed a buckled carbon sheet with tetrarings "T-graphene" and thought it as a 2D Dirac material[116]. Although T-graphene was demonstrated to be metallic soon later[117], it drives people to notice the 2D carbon allotropes with multi-member rings. In 2014, Xu et al. designed a series of 2D carbon allotropes by reconstructing graphene[23]. Moreover, three stable rectangular systems named as S-graphene (Fig. 6a), D-graphene (Fig. 6b), and E-graphene (Fig. 6c) were predicted to have Dirac cones[23]. S-graphene consists of eight C atoms in a unit cell and contains four- and six-member rings. It has two self-doping Dirac cones along **M**-**X** and **Y**-$\Gamma$ lines separately. D-graphene is composed of sp and sp$^2$ C atoms while E-graphene composed of sp$^3$ and sp$^2$ C atoms. Both of them have distorted Dirac cones, but the Dirac points are on different positions. More carbon sheets with different rings and hybridized forms have also been suggested to be Dirac materials[23].

Boron—the neighbor of carbon in the periodic table—is a fascinating element because of its chemical and structural complexity. The energy landscape of boron clusters is glasslike, being in sharp contrast to carbon and boron nitride systems[118]. Boron might be the second element that can possess freestanding flat monolayer structures, evidenced by the experimental synthesis of single-walled and multiwalled boron nanotubes and the stability calculation of boron sheet[119]. Boron sheets are usually composed of triangular and hexagonal motifs arising from the competition between two- and three-center bonding[120]. Recently, a novel 2D boron sheet (*Pmmn* boron, Fig. 6d) has been predicted to have distorted Dirac cones[24]. It has eight atoms in a rectangular cell with the lattice constants of 4.52 and 3.26 Å. The B atoms form a buckled triangular layer that can be classified into two sublattices (buckled chains and hexagons, illustrated in Fig. 6d by different colors). The predicted Dirac points locate at the **Γ-X** line, and the hybrid of in-plane ($p_x$ orbitals from the buckled boron chains) and out-of-plane states ($p_z$ orbitals from the buckled irregular boron hexagons) is responsible for the emergence of Dirac cone[24].

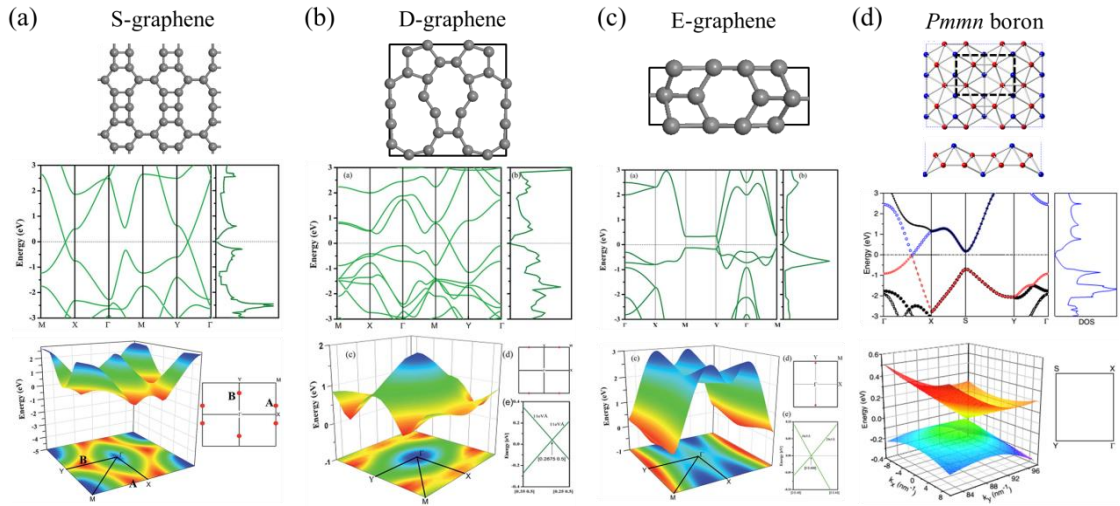

Figure 6. Atomic and electronic structures of (a) S-graphene[23] (Copyright 2014 Royal Society of Chemistry), (b) D-graphene[23] (Copyright 2014 Royal Society of Chemistry), (c) E-graphene[23] (Copyright 2014 Royal Society of Chemistry), and (d) *Pmmn* boron[24] (Copyright 2014 American Physical Society).

**Transition metal oxides:** $(VO_2)_n/(TiO_2)_m$ **and** $(CrO_2)_n/(TiO_2)_m$

Dirac cones also exist in multilayer nanostructures of transition metal oxides. The system $(VO_2)_n/(TiO_2)_m$, formed by $n$ $VO_2$ monolayers and $m$ $TiO_2$ monolayers grown along the rutile (001) direction, was predicted to possess Dirac cones for $n = 3$ and 4, where the spectrum in the vicinity of the Fermi level is dominated by V 3$d$ orbitals[30]. The in-plane cell of this pseudo-2D system is square, but it does not hold the full square symmetry and thus escape the restriction of the full square symmetry on the existence of Dirac cones[121]. Remarkably, although the dispersion is linear along the diagonal (**Γ-M**) direction, it is quadratic perpendicular to the diagonal[30, 122]:

$$E(\mathbf{k}) = \pm\sqrt{(v_F k_1)^2 + (k_2^2/2m)^2} \tag{13}$$

where $\upsilon_F$ and $m$ are the Fermi velocity and effective mass, respectively. It is different from the usual Dirac cones as that in graphene, and was thus referred to as semi-Dirac cones[30]. The strongly anisotropic dispersion is expected to give rise to peculiar transport and thermodynamic properties. Similar semi-Dirac point has also been found in a photonic crystal consisting of a square array of elliptical dielectric cylinders[123] and a microwave graphene analogue composed of coupled dielectric cylindrical resonators[124]. Recently, a close inspection has revealed that the semi-Dirac cones in $(VO_2)_n/(TiO_2)_m$ are formed by merging three usual Dirac cones at each semi-Dirac point, which results in a nontrivial Chern number in this system[1]. In addition, usual Dirac cones have been predicted in $(CrO_2)_n/(TiO_2)_m$ for $n = 4$[125].

**Organic and organometallic crystals**

Another class of materials which affords Dirac cones is organic conductors. A realized example is the quasi-2D organic conductor α-(BEDT-TTF)$_2$I$_3$ under high pressure[28, 126]. α-(BEDT-TTF)$_2$I$_3$ is composed of conductive layers of BEDT-TTF molecules and insulating layers of $I_3^-$ anions, and the in-plane cell is rectangular[126]. According to the band calculation, α-(BEDT-TTF)$_2$I$_3$ possesses two Dirac points at the Fermi level, which, in contrast to the case of graphene, locate at non-symmetric positions in the **k**-space[28]. In addition, the Dirac cones are heavily tilted and the Fermi velocity along **k** is not equal to –**k**, i.e., the inversion symmetry of the Dirac cones is lost. Experimentally, the zero-mode Landau level expected to appear at the Dirac points has been successfully detected in the magnetoresistance measurement[127].

The molecular interactions in organic crystals such as α-(BEDT-TTF)$_2$I$_3$ are much weaker than the interactions in atomic crystals such as graphene and silicene. As a result, the energy scale in organic crystals is small and their applications at the room temperature are limited. To remedy this shortcoming, metal atoms can be introduced to glue organic compounds. Organometallic crystals Pb$_2$(C$_6$H$_4$)$_3$, Ni$_2$(C$_6$H$_4$)$_3$ and Co$_2$(C$_6$H$_4$)$_3$ with a hexagonal lattice have recently been proposed to possess Dirac cones[22, 26]. Due to the incorporation of the magnetic Ni and Co atoms, Ni$_2$(C$_6$H$_4$)$_3$ and Co$_2$(C$_6$H$_4$)$_3$ are half-metallic, which may extend the applications of 2D Dirac materials in spintronics[22].

**Systems with a pseudospin of $S = 1$**

Dirac cones in most Dirac materials can be described by Dirac-Weyl equation with pseudospin $S = 1/2$ as Eq. (5). However, Dirac cones with unusual spin are also possible. Wang et al. have constructed various carbon allotropes with square symmetry by enumerating the carbon atoms in a unit cell up to 12[21]. One of them, SG-10b (Fig. 7a), was predicted to possess Dirac cones with a pseuspin of $S = 1$ in a TB calculation[21]. Two cone-like bands and one flat band contact at the **M** point at the Fermi level (Fig. 7a), which can be described by a Dirac-Weyl Hamitonian:

$$H = \upsilon_F \mathbf{q} \cdot S \tag{14}$$

where $S$ is a pseudospin of the quantum number 1 with three eigenvalues −1, 0 and +1. The Massless Dirac-Weyl fermions with a pseudospin $S = 1$ have interesting properties that are different from $S = 1/2$. The Berry's phase enclosing the Dirac point of $S = 1$ vanishes, so the Dirac point of $S = 1$ can exist solely, unlike those of $S = 1/2$ that appears in pairs. The enlarged pseudospin also leads to an enhanced Klein tunneling where the barrier is transparent for all incident angles[128]. In addition, the presence of the flat band under $S = 1$ serves as a good starting point to study correlated electron systems[129].

---

[1] Huang, H. Q.; et-al., Chern insulator in TiO2/VO2 nanostructures. preprint.

SG-10b is not the only atomic crystal predicted to possess Dirac cones with $S = 1$. The odd-walled hexagonal graphene antidote lattices (GALs) also possess massless Dirac fermions with $S = 1$ at the Fermi level determined by first-principles calculations (Fig. 7b)[20-21]. Another example is a $MoS_2$ allotrope[25]. Although normal monolayer $MoS_2$ with a hexagonal lattice ($h$-$MoS_2$) is a direct-gap semiconductor[130], the square $MoS_2$ sheet (so-$MoS_2$) has been shown to possess massless Dirac fermions from $d$ electrons (Fig. 7c)[25]. The only Dirac point locates at the **Γ** point, so it can be recognized to be $S = 1$. Recently, a modified TB model based on $T_3$-lattice has provided a continuous interpolation between the fermions with $S = 1/2$ and the fermions with $S = 1$[131]. More 2D Dirac systems with $S = 1$ can be expected.

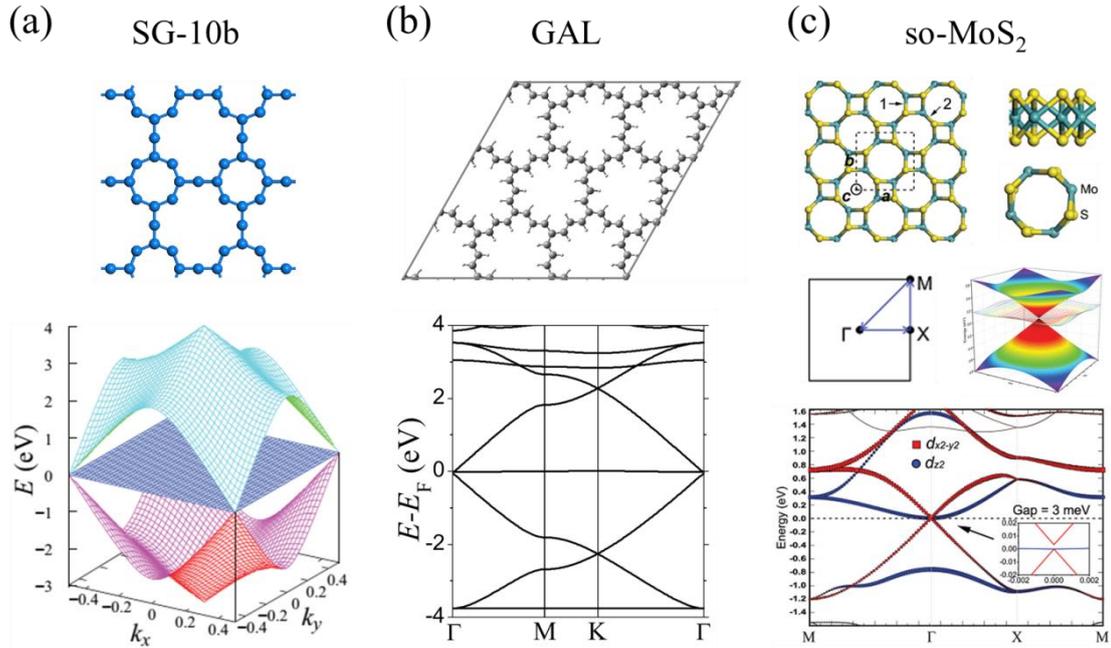

Figure 7. Atomic and band structures for (a) SG-10b[21] (Copyright 2013 AIP Publishing LLC), (b) a single-walled hexagonal GALs[21] (Copyright 2013 AIP Publishing LLC), and (c) so-$MoS_2$[25]

(Copyright 2014 American Physical Society).

**Artificial lattices: electron gases and ultracold atoms**

The extraordinary properties of graphene due to the existence of Dirac cones have stimulated the search of the "quantum simulators" of Dirac cones in artificial systems with distinct mechanisms, e.g., 2D electron gases, ultracold atoms and photonic crystals[132].

Ultracold atoms trapped in optical lattices serve to mimic condensed matter phenomena very well. Realizing Dirac cones with cold atoms loaded in a honeycomb optical lattice was first proposed in 2007[29]. Cold atom systems with orbital degrees of freedom can even afford new features which are impossible to realize in graphene, e.g., $p_{x,y}$-orbital counterpart of graphene with flat bands and Wigner crystallization[133-134]. Cold atoms with other structures of symmetries can also afford Dirac cones. For example, atoms trapped in a $T_3$ lattice or a line-centered-square lattice behave as the massless Dirac fermions with pseudospin $S = 1$[135-136]. In 2012, the experimental creation of Dirac cones at **K** and **K**' was reported in an ultracold gas of $^{40}$K atoms trapped in a 2D honeycomb optical potential lattice of interfering laser beams[137]. Moreover, with changing the lattice parameters, the positions of the two Dirac points inside the Brillouin zone move accordingly, and finally merge and annihilate each other when the parameters exceed a critical limit—a situation that is extremely challenging to observe in solids.

2D electron gases are gases of electrons free to move in two dimensions but tightly confined in the third. They are usually found in nanostructures made from semiconductors. A theoretical analysis on the electronic structure of 2D electron gases under an external periodic potential of hexagonal symmetry was conducted in 2009, which revealed that Dirac cones are generated at **K** and **K**' as those in graphene[138-139]. The required parameters were estimated to be achievable under laboratory conditions for semiconductors. In 2012, by combining both experimental and theoretical studies, it was revealed that the main obstacle preventing the realization of Dirac-like physics in 2D electron gases for semiconductors is related to the interplay between the electron density and the lattice constant[140]. In the same year, a different scheme was developed to successfully achieve Dirac cones at a Cu surface[27]. Cu(111) surface is characterized by its nearly free electron gases with very long coherence lengths. Adsorbed CO molecules on clean Cu(111) were assembled into a lattice by positioning the molecules individually using the tip of a scanning tunneling microscope (STM), which exert periodic potential over the free electron gases. With CO molecules in a hexagonal lattice, the existence of linearly dispersing, massless quasi-particles and a density of states characteristic of graphene have been clearly shown in the experiments[27]. More importantly, the CO lattice sites can be arbitrarily tuned globally or locally to mimic various effects, providing a versatile means of synthesizing exotic electronic phases in condensed matter.

Photonic crystals are periodic optical nanostructures that affect the motion of photons in a way similar to how ionic lattices affect electrons in solids. Photonic crystals offer a distinctive route to obtain dispersion relations with characteristic Dirac cones[132, 141-142]. However, photons are boson, being markedly different from the other systems discussed in this review which are fermion, and photons rarely exist in the equilibrium state. Therefore, advances of photonic crystals with Dirac cones will not be addressed here. Readers are directed to reference[132] which provides an excellent review in this field. New 2D Dirac systems keep springing up[143], but they are rather rare compared to the numerous 2D materials.

**Why are Dirac materials so rare?**

**Moving and merging of Dirac points**

Dirac cones are rather robust under perturbation. For example, when a uniaxial or shear strain is applied, the band structure of graphene keeps gapless and the Dirac point moves to a new **k** location near the original one[51-52]. A TB analysis on the honeycomb lattice with the nearest neighboring interactions indicated that Dirac cones exist in the system as long as three hopping integrals ($t_1$, $t_2$, $t_3$) can form a triangle[144]:

$$|t_2 - t_1| \leq |t_3| \leq |t_2 + t_1|. \tag{15}$$

For a general 2D system that contains two atoms of the same species in each unit cell, the criterion for the existence of Dirac cones becomes

$$|t_1^2 + t_2^2 - t_3^2 - t_4^2| \leq |2t_1 t_2 - 2t_3 t_4|, \tag{16}$$

and the locations of Dirac cones are given explicitly in a function of four hopping integrals ($t_1$, $t_2$, $t_3$, $t_4$)[121]. The effect of strain on Dirac cones can be well explained in terms of the variation of hopping integrals[52].

When there is no external magnetic field, the system is invariant under the time-reversal operation, where a Dirac point at $\mathbf{k}_0$ is always accompanied by its pair at $-\mathbf{k}_0$, e.g., **K** and **K'** for graphene. The Berry phase around a Dirac point is either $+\pi$ or $-\pi$, and the Berry phase around a Dirac point at $\mathbf{k}_0$ is always opposite to that around its pair at $-\mathbf{k}_0$. When two Dirac points with opposite Berry phases move in the **k** space under any perturbation and arrive at the same point, they merge and their Berry phases annihilate each other[121, 145-146]. At the merging point, the energy dispersion is semi-Dirac which remains linear in one direction but becomes quadratic in the other. Upon further perturbation, a gap is induced and the Dirac points disappear. If there is only a single pair of Dirac points in the system, they can be merged only at high symmetric **k**-points. For example, the Dirac point observed near **X** in 6,6,12-graphyne is actually formed by merging two Dirac points[18-19]. An in-plane AC electric field can also drive the moving and merging of Dirac cones[147]. On the other hand, the opposite processes are also possible, i.e., the energy gap may disappear and a pair of Dirac points emerges at high symmetric **k**-points. For example, DFT calculation showed that a uniaxial strain of 9% induced a pair of Dirac points at **Γ** in graphdiyne[106].

To achieve Dirac cone merging or emerging is not an easy task in experiments. The critical strain required to annihilate Dirac cones in graphene is as high as 24% due to the well separated Dirac points originally at **K** and **K**'[51]. Such a large strain is experimentally impractical. A possible way to reduce the modulation difficulty is to replace some single bonds in graphene with acetylenic linkages (−C≡C−), i.e., to adopt graphynes or graphdiynes. For example, one pair of Dirac points in 6,6,12-graphyne locates very close to the high symmetric point **X**, which is possible to annihilate at small strain. Unfortunately, the controlled preparation of graphynes and graphdiynes is difficult by itself. At present, Dirac cone merging is achieved only in artificial honeycomb lattices where parameters are much more adjustable. By patterning CO molecules on clean Cu(111), the hexagonal potential lattice of electron gases was effectively modulated to demonstrate a transition from massless to massive Dirac fermions in the system[27]. In an ultracold gas of $^{40}$K atoms trapped in a 2D honeycomb optical potential lattice, the merging and annihilation of two Dirac points were clearly recorded when the lattice anisotropy exceeds a critical limit[137].

Interestingly, more than two Dirac points can also merge together. When three Dirac points merge, their Berry phases cannot completely annihilate each other and thus possess a nontrivial topology. Quasi-2D $(VO_2)_3/(TiO_2)_m$ possesses semi-Dirac cones as described by Eq. (13) in its band structure, but the Berry phase around the semi-Dirac points is ±π and the Chern numbers are nonzero, which were actually caused by the merging of three Dirac points[2]. When four Dirac points merge together, the energy dispersion becomes parabolic in any direction near the merging points. This is what happens in bilayer graphene. Its nature of four-Dirac-point-merging can be clearly demonstrated by exerting the next-nearest-neighbor interlayer hopping: each merging point will split into four conventional Dirac points with linear dispersions[12]. The Berry phase around the merging points in bilayer graphene is +2π (–2π), suggesting that they are merged by three +π (–π) and one –π (+π). An illustration for the various situations is given in Fig. 8. Although the merging of two and three Dirac cones both lead to a semi-Dirac cones near which the energy dispersion is linear in one principal axis and quadratic in the other, the band structure in the merging of three Dirac cones is more twisted, appearing in a banana shape[3].

---

[2] Huang, H. Q.; et-al., Chern insulator in TiO2/VO2 nanostructures. preprint.
[3] Huang, H. Q.; et-al., Chern insulator in TiO2/VO2 nanostructures. preprint.

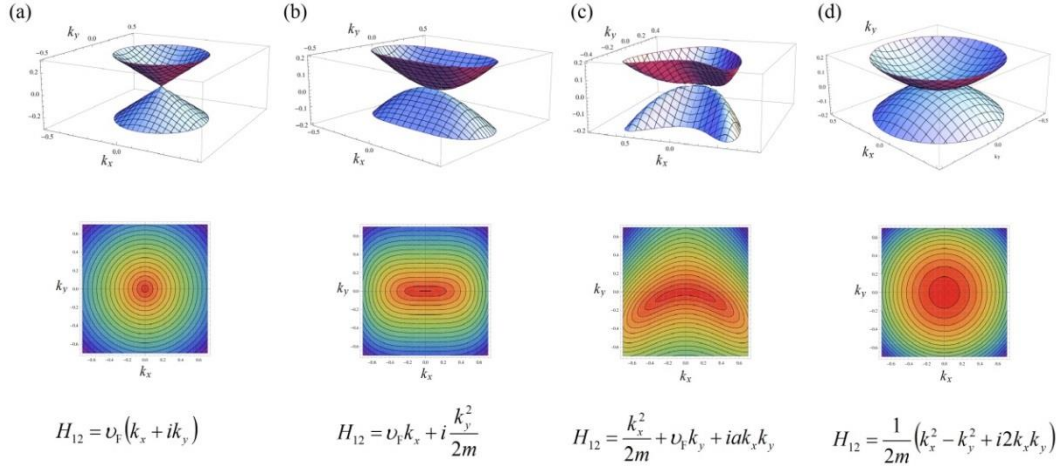

Figure 8. The merging of (a) one, (b) two, (c) three, and (d) four Dirac points[4]. Three dimensional band structure and 2D energy contour are shown at top and bottom panels, respectively. A simplified 2×2 Hamiltonian of $\begin{bmatrix} 0 & H_{12} \\ \text{c.c.} & 0 \end{bmatrix}$ is adopted with the expression of $H_{12}$ given at the bottom for each case.

## Existent conditions of Dirac cones

Dirac points are not just energy band crossings but also singularities in the spectrum of Hamiltonians[148]. Dirac cones are usually topologically protected and have inherent connection to the $Z_2$ quantization defined in terms of Berry phases[149-150]. However, why Dirac cones are not ubiquitous but rather rare in 2D materials?

Briefly speaking, the rarity of Dirac cones in 2D systems can be explained with the von Neumann-Wigner theorem[151-152]. Considering a general two-band system whose Hamiltonian is given as:

$$H(\mathbf{k}) = \begin{bmatrix} H_{11}(\mathbf{k}) & H_{12}(\mathbf{k}) \\ \text{c.c.} & H_{22}(\mathbf{k}) \end{bmatrix} \tag{17}$$

and we obtain

$$E_{\pm}(\mathbf{k}) = \frac{H_{11}(\mathbf{k}) + H_{22}(\mathbf{k})}{2} \pm \sqrt{\left(\frac{H_{11}(\mathbf{k}) - H_{22}(\mathbf{k})}{2}\right)^2 + |H_{12}(\mathbf{k})|^2} \tag{18}$$

The conditions to determine the Dirac point are thus

---

[4] Huang, H. Q.; et-al., Chern insulator in TiO2/VO2 nanostructures. preprint.

$$\begin{cases} H_{11}(\mathbf{k}) - H_{22}(\mathbf{k}) = 0 \\ \mathrm{Re}\left[H_{12}(\mathbf{k})\right] = 0 \\ \mathrm{Im}\left[H_{12}(\mathbf{k})\right] = 0 \end{cases} \quad (19)$$

where $\mathrm{Re}\left[H_{12}(\mathbf{k})\right]$ and $\mathrm{Im}\left[H_{12}(\mathbf{k})\right]$ denote the real and imaginary part of the complex number $H_{12}$. The three conditions in Eq. (19) must be simultaneously fulfilled to have an energy degeneracy, which is known as the von Neumann-Wigner theorem[151]. In 2D systems, we have two variables ($k_x$, $k_y$) to be solved from the above three equations. Because the number of variables is less than the number of equations, the problem is overdetermined and we usually get no solution. This is the main reason why Dirac cones are so rare.

To make the existence of Dirac points possible, a constraint on $H(\mathbf{k})$ is required, which would reduce the number of conditions in Eq. (19) by one[152]. The required constraint is usually attributed to the symmetries of the system. A representative example is the space-time inversion symmetry[152-153]. The invariance under the time reversal, $T$, gives

$$T : H(\mathbf{k}) = H^*(-\mathbf{k}) \quad (20)$$

and the spatial inversion gives

$$I : H(\mathbf{k}) = \sigma_x H(-\mathbf{k}) \sigma_x \quad (21)$$

where $\sigma_x$ is a Pauli matrix. Under a combination of $T$ and $I$,

$$TI : H(\mathbf{k}) = \sigma_x H^*(\mathbf{k}) \sigma_x \quad (22)$$

which leads to $H_{11}(\mathbf{k}) = H_{22}(\mathbf{k})$ and cancels the first equation in Eq. (19), making the existence of Dirac cones possible. Most Dirac materials, if not all, are actually protected from this kind of space-time inversion symmetry.

Not all symmetries are effective in providing constraint on $H(\mathbf{k})$. Generally, a symmetry operation would relate the Hamiltonian at a point $\mathbf{k}$ to that at another point $\mathbf{k}'$, e.g., Eqs. (20, 21). Although they are useful properties, what we need here is a constraint on the form of $H(\mathbf{k})$. Therefore, to provide a constraint, the $\mathbf{k}$-points should be kept unchanged after the symmetry operation (invariant $\mathbf{k}$-point). The combination of $T$ and $I$ meets such a requirement for a general $\mathbf{k}$ point as shown in Eq. (22), but the separate $T$ or $I$ does not. Similarly, a sole rotation cannot provide an effective constraint either and is not essential for the existence of Dirac cones. For example, when a shear strain is applied upon graphene, although the rotation symmetries are destroyed, Dirac cones exist in the system before the strain exceeds a large threshold[51-52].

Although the achieved constraint from the system symmetries makes the existence of Dirac cones possible, it does not guarantee it. The major reason is that $k_x$ and $k_y$ are real number and appear in $H(\mathbf{k})$ in the form of a sine or cosine function. The solution of Eq. (19) may be still absent for certain parameter values even with the constraint[121]. Eq. (15) and Eq. (16) are two examples of the required parameters to guarantee Dirac cones. Another excellent example comes from a comparison between $β$- and $γ$-graphynes. $β$- and $γ$-graphynes share completely identical symmetries, but $β$-graphyne is semimetal with Dirac cone while $γ$-graphyne is semiconductor. A detailed analysis revealed that the hopping integrals in $β$-and $γ$-graphynes are responsible for their existence/absence of Dirac cones[19]. Hopping integrals are affected by atomic geometry such as bond length, which is further related to the crystal lattice. A simplified analysis on a general 2D system, which contains two atoms of the same species in each unit cell, demonstrated that a hexagonal cell is the most favorable for the existence of Dirac cones, and the favorableness is gradually diminished when the cell evolves into a square one (Fig. 9)[121]. This is consistent with the fact that many Dirac materials are observed in hexagonal lattice.

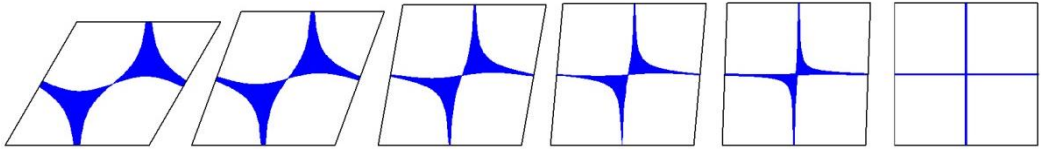

Figure 9. Hexagonal cell is favorable for the existence of Dirac cones. A system containing two atoms in each unit cell is considered with the first atom located at corners. Location regions of the second atom, which enable the existence of Dirac cones, are marked by the filled areas[121].

To observe the novel properties particular to the Dirac cones, the Fermi level should lie at the Dirac points and no other bands go through the Fermi level. There should not have any other bands than Dirac points overlap at the Fermi level, otherwise the low-energy excitation would be dominated by the carriers in the hole and electron pockets caused by the overlap. For example, although quasi-2D Dirac cones were predicted in $LaAgBi_2$[154], they may be difficult to detect experimentally. The band overlap can be avoided when the diagonal elements in Eq. (17), $H_{11}(\mathbf{k})$ and $H_{22}(\mathbf{k})$, are $\mathbf{k}$-independent. This condition is automatically satisfied in a bipartite system under a nearest-neighbor approach. Actually, all Dirac materials listed in Fig. 1 are bipartite except *Pmmn* boron. It is noted that being bipartite is not sufficient condition to possess Dirac cones. For example, the single-layer hexagonal BN (*h*-BN) sheet has a bipartite lattice as graphene, but its two $\mathbf{k}$-independent $H_{11}$ and $H_{22}$ are not equal to each other due to the different onsite energies of B and N atoms, so it is semiconductor with a large energy gap[155].

Symmetries are usually needed to produce Dirac cones, but too high symmetries may be disadvantageous. If there is too much constraint on $H(\mathbf{k})$, the number of independent equations may be smaller than the number of independent variables, and thus the obtained solutions are not discrete points, but a continuous line in the $\mathbf{k}$-space. In such a case, the carriers near the Fermi level cannot be described by massless Dirac equation, and the corresponding novel properties are absent. For 2D carbon allotropes with full square symmetry (wallpaper group *p4mm*), many systems have band contact at the Fermi level appearing in lines[21]. Therefore, they are metal, and the absence of Dirac cones in them is not caused by the absence of contacts between valence and conductance bands, but caused by the existence of too much contacts. The only carbon allotrope with full square symmetry recognized to possess Dirac cone, SG-10b, has three bands contacting at the Dirac point, but not two, and cannot be described by the Eq. (17)[21]. A study on a general 2D atomic crystal containing two atoms in each unit cell and a 2D electron gas under a periodic muffin-tin potential also suggested that conventional Dirac cones with $S = 1/2$ (such as those observed in graphene) are difficult, if not impossible, to be achieved under full square symmetry[121]. As a support for this opinion, it is noted that although the Dirac material $(VO_2)_n/(TiO_2)_m$ and $(CrO_2)_n/(TiO_2)_m$ have a pseudo-2D square lattice, they do not possess full square symmetry.

Based on the analyses given above, we can have a brief summary on why Dirac cones are so rare. The central origin is the von Neumann-Wigner theorem: the number of variables ($k_x$ and $k_y$) is usually less than the number of equations to determine the Dirac points. In more details, to achieve Dirac materials, at least three conditions are required: (1) Symmetry. Specific symmetries are required to reduce the number of equations to be solved. The **k**-points should be unchanged after the symmetry operation (invariant **k**-point). Too low or too high symmetries are both disadvantageous. (2) Parameters. Even when the number of equations is equal to the number of variables, the solution is not necessarily exist since the variables ($k_x$ and $k_y$) are real numbers and appear in the equations in the form of a sine or cosine function. Therefore, proper parameters are required. This is usually described as a phase diagram in the parameter space. (3) Fermi level and band overlap. The Fermi level should lie at the Dirac points and there should not have any other band than Dirac points overlap at the Fermi level. The rigorous conditions may explain the scarcity of Dirac materials. It is noted that the energy dispersion near the Dirac point solved from the above conditions is linear in most cases, because a quadratic dispersion needs more constraints to satisfy $dE/d\mathbf{k} = 0$ at the crossing point. The understanding on the conditions for Dirac materials also provides a guide to search/design new Dirac materials in the future.

## Structural features of the known 2D Dirac materials

After discussing the conditions for Dirac cones to exist, we now briefly analyze the common structural features of the known atomic Dirac materials to see how they are beneficial for the existence of Dirac cones.

First, most Dirac materials have spatial inversion symmetry. As demonstrated above, in combination with time reversal, spatial inversion provides an effective constraint on $H(\mathbf{k})$ to reduce the number of equation. The exceptions are transition metal oxides of $(VO_2)_n/(TiO_2)_m$ and $(CrO_2)_n/(TiO_2)_m$. Their Dirac points are protected by mirror symmetry. As a result, the Dirac points locate along the **Γ-M** lines which are invariant under mirror operation.

Secondly, all atomic Dirac materials have even number of atoms in a unit cell. Many of them are bipartite and composed of only one element (mainly belongs to group IVA). The bipartite feature forces the diagonal Hamiltonian elements to be constants, and thus is favorable not only in reducing the equation number but also in avoiding band overlap at the Fermi level. Single element, on the other hand, is helpful for the on-site energies at different sites to be equal. The different capacity of graphene and *h*-BN can be explained in terms of relation between the element number and the on-site energy.

We can transform the materials into binary systems by introducing the concept of superatoms. In detail, we replace half atoms in a unit cell of each system in Fig. 1 except $(VO_2)_n/(TiO_2)_m$ by one superatom and the other half by another superatom (Fig. 10c), and redraw the structures. Then graphene, α-graphyne, silicene, germanene, CO on Cu(111) and $Pb_2(C_6H_4)_3$ can be transformed to a graphene-like structure (Fig. 10a), and the 6,6,12-graphyne, 14,14,18-graphyne, S-graphenen, SG-10b, *Pmmn* boron, and so-$MoS_2$ can be transformed to a rectangular or square structure constituted by two superatoms in spatial inversion (Fig. 10b).

Lastly, hexagonal honeycomb structure is common in atomic Dirac materials. This feature may relate to the requirement of parameters. As illustrated in Fig. 9, hexagonal cell is more favorable in providing appropriate parameters for the existence of Dirac cones. We take an effective bond method to analyze the topological structure of the systems. Assume that C-C, Si-Si, Ge-Ge, and C-C≡C-C bonds can be expressed as one effective bond, and then the elementary substances in Fig.1 have the same honeycomb topology (Fig. 10d). For *Pmmn* boron, since it has two sublattices which contribute the valence and conduction band separately[24], we connected the atoms in the same sublattice and found that two honeycomb structures are formed (Fig.10e).

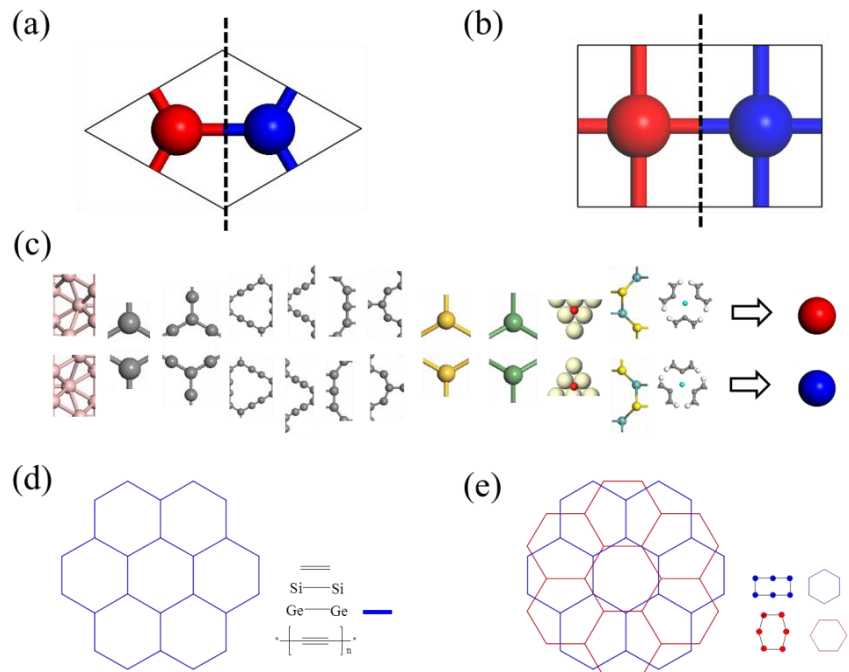

Figure 10. (a) honeycomb and (b) rectangular structure for 2D Dirac materials by introducing (c) the concept of superatoms. Topological structures for (d) graphene, silicene, germanene, graphynes and (e) *Pmmn* boron.

## Conclusions

Dirac cones are not only the linear energy dispersion around discrete points but also singularities in the spectrum of Hamiltonians and are topologically protected. Dirac-cone structures give materials unusual properties and promising prospects in both fundamental and applied research. However, the probability of Dirac cones in 2D systems is rather low. Herein, we have systematically surveyed the known 2D Dirac materials and discussed how Dirac cones emerge and merge in the system. Rigorous conditions on symmetry and parameters are required to achieve Dirac cones in 2D systems, which provide an explanation for the rarity of Dirac materials. Looking forward, we believe that more and more 2D Dirac materials will be discovered, and a thorough understanding on the existing conditions of Dirac cones is greatly helpful in seeking/designing new systems.

**Acknowledgements**

We are grateful to Wenhui Duan, Huaqing Huang, Shuqing Zhang and Zhenzhu Li for discussions. This work was supported by the National Natural Science Foundation of China (Grant No. 21373015).